\documentclass[preprint]{ptephy_v1}

\preprintnumber{NITEP 139} 

\usepackage{ulem,bm}

\begin{document}

\title{Incomplete absorption reactions at high energy}

\author[1]{Katsuhito Makiguchi}

\author[2,3,1*]{Wataru Horiuchi}

\affil{Department of Physics,
  Hokkaido University, Sapporo 060-0810, Japan}
\affil[2]{Department of Physics, Osaka Metropolitan University, Osaka 558-8585, Japan   }
\affil[3]{Nambu Yoichiro Institute of Theoretical and Experimental Physics (NITEP), Osaka Metropolitan University, Osaka 558-8585, Japan
\email{whoriuchi@omu.ac.jp}}

\begin{abstract}
The total reaction cross section of high-energy nucleus-nucleus collision
reflects the nuclear density profiles of the colliding nuclei
and has been a standard tool to investigate
the size properties of short-lived unstable nuclei.
This basis relies on the assumption that the nucleus-nucleus
collision is strongly absorptive in the sense of an optical model.
However, this property does not hold completely, while incomplete
absorption occurs when an overlap density of two colliding nuclei
is low enough. In this paper, we propose a way to quantify this incompleteness,
that is, the ``blackness'' of the total reaction cross section.
A significance of this quantification is drawn
by taking an example of the total reaction cross sections
of proton-rich C isotopes.
\end{abstract}

\subjectindex{xxxx, xxx}

\maketitle

\section{Introduction}

Total reaction cross sections of nucleus-nucleus collision at
a few tens to thousand MeV have been utilized to extract the nuclear
size properties of short-lived unstable nuclei.
These intensive measurements have revealed,
a two-neutron halo nucleus~\cite{Tanihata85}, the development
of neutron skin~\cite{Suzuki95},
and strong deformations of unstable nuclei~\cite{Takechi12,Takechi14}. 
See, e.g., Ref.~\cite{Tanihata13} for reviews of those advancements.
Such studies have been extended to F isotopes near dripline nucleus
$^{29}$F exhibiting two-neutron halo structure~\cite{Bagchi20},
and Ca isotopes showing a sudden core swelling phenomenon~\cite{Tanaka20}.
The total reaction cross section measurements for the unstable nuclei
are always made
in the inverse kinematics using a well-known and stable target nucleus,
typically, carbon and proton targets.
As those target nuclei offer different sensitivity
to the nuclear density profile of the projectile nucleus,
quantifying the sensitivity opens up the possibility of
determining the selected density profile in exotic neutron-rich nuclei.

Since the beam energy is high and the nuclear interaction is short,
the total reaction cross section at the high incident energies
can be approximated well by a black-sphere (BS) model~\cite{Kohama04,Kohama05,Kohama16},
and hence the cross section is roughly proportional to $\pi (R_P+R_T)^2$,
where $R_P$ and $R_T$ are the nuclear radii of the projectile
and target nuclei, respectively.
This is a basis that the nuclear radius can be extracted
from the total reaction cross section.
However, because the nuclear reaction is not a perfect black sphere,
the information more than the nuclear radius, i.e.,
the density profile near the nuclear surface, can be extracted 
~\cite{Kohama16,Hatakeyama18,Makiguchi20}.
As it may be related to that incompleteness, in Ref.~\cite{Kaki17},
an interesting observation was made in the nucleus-proton total reaction cross
sections for proton-rich C isotopes:
Despite that, the nuclear matter radius is enhanced,
the total reaction cross section is reduced.
This implies the deviation from the BS picture
of the high-energy nuclear collision.
The purpose of this paper is to quantify
the ``blackness'' of high-energy nucleus-nucleus reactions
and discuss the effect of its incompleteness
on the total reaction cross sections. 

The paper is organized as follows.
To quantify the nuclear blackness we employ a microscopic
high-energy reaction theory, the Glauber model.
In Sec.~\ref{method.sec}, we briefly describe
the basic inputs of the Glauber model.
Section~\ref{results.sec} is devoted
to discuss our results.
Section~\ref{blackness.sec} quantifies
the blackness of the high-energy nucleus-nucleus collisions.
We conveniently divide the total reaction cross sections into
the complete and incomplete absorption parts 
based on the reaction probabilities
obtained by the present microscopic reaction model.
In Sec.~\ref{carbon.sec},
the characteristic behavior of the total reaction cross sections
as a function of the neutron number is discussed by
taking an example of proton-rich C isotopes,
in which the incomplete absorption part is dominant.
Conclusion is made in Sec.~\ref{conclusion.sec}.

\section{Method}
\label{method.sec}

To describe the transparency in the high-energy nucleus-nucleus collsion,
we employ a microscopic high-energy nuclear theory,
the Glauber model~\cite{Glauber}.
The total reaction cross section is calculated
by integrating the reaction probability
\begin{align}
  P(\bm{b})=1-|e^{i\chi(\bm{b})}|^2,
\label{reacprob.eq}
\end{align}
over the impact parameter vector $\bm{b}$ as
\begin{align}
  \sigma_R=\int d\bm{b}\, P(\bm{b}).
\end{align}
The reaction probability describes how the incoming flux is absorbed
by the target nuclei. Note that
\begin{align}
  P(\bm{b})=
 \begin{cases}
      1 & (b\leq R)\\
      0 & (b>R)
 \end{cases}
 \label{BS.eq}
\end{align}
holds for the BS or complete absorption model
with a sharp nuclear radius $R$.
In the Glauber model,
it is vital to evaluate the optical phase-shift function 
\begin{align}
e^{i\chi(\bm{b})}=\left<\Psi_0^P\Psi_0^T\right|\prod_{j\in{\rm P}}^{A_P}
\prod_{k\in{\rm T}}^{A_T}[1-\Gamma_{NN}(\bm{s}_j^P-\bm{s}_k^T+\bm{b})]\left|\Psi^P_0\Psi_0^T\right>,
\label{PT.psf}
\end{align}
where $\bm{s}_j^P$ ($\bm{s}_j^{T}$)
is the two-dimensional single-particle coordinate
of the $j$th nucleon in the projectile (target) nucleus
perpendicular to the beam direction $z$,
$\Psi_0^P$ ($\Psi_0^T$) is the ground-state wave function
of the projectile (target) nucleus,
and $\Gamma_{NN}$ is the profile function,
which describes the nucleon-nucleon ($NN$) collision
and is usually parametrized as
\begin{align}
\Gamma_{NN}(\bm{b})=\frac{1-i\alpha_{NN}}{4\pi\beta_{NN}}
\sigma_{NN}^{\rm tot}\exp\left[-\frac{\bm{b}^2}{2\beta_{NN}}\right],
\label{profn.eq}
\end{align}
where $\alpha_{NN}$ is the ratio of the real to the imaginary part of the $NN$
scattering amplitude of forward angle, $\beta_{NN}$ is
the slope parameter of the $NN$ elastic scattering differential cross section,
and $\sigma_{NN}^{\rm tot}$ is the total cross section of the $NN$ scattering. 
To describe the isospin dependence appropriately, especially for
the proton scattering, we take the neutron-proton ($np$) profile function
being different from the $pp$ one. 
The $nn$ profile function is taken the same as that for $pp$.
These parameter sets are tabulated
in Ref.~\cite{Ibrahim08} for a wide energy range.

The evaluation of the phase-shift function of Eq. (\ref{PT.psf})
involves multi-dimensional integration, which
may be done with a Monte Carlo technique~\cite{Varga02, Nagahisa18}.
However, in general, it is demanding. Thus, the optical phase-shift function is
usually approximated by taking the leading order of the cumulant expansion as
\begin{align}
  &i\chi(\bm{b})=
  -\iint d\bm{r}^Pd\bm{r}^T\,\rho^P(\bm{r}^P) \rho^T(\bm{r}^T)
  \Gamma_{NN}(\bm{s}^P-\bm{s}^T+\bm{b}),
\end{align}
where $\bm{r}^{P(T)}=(\bm{s}^{P(T)},z^{P(T)})$
denotes the two-dimensional coordinate
of the projectile (target) nucleus perpendicular
to the beam direction $z^{P(T)}$,
and $\rho^P(\bm{r}^{P})$ ($\rho^T(\bm{r}^T)$)
is the density distribution of the projectile (target) nucleus.
This expression is called optical limit approximation (OLA)
and works well for elastic and total reaction
cross sections of nucleus-proton scattering where
the multiple scattering effect
is negligeble~\cite{Varga02, Ibrahim09, Hatakeyama14,Hatakeyama15, Nagahisa18, Hatakeyama19}.
For the nucleus-nucleus collision,
the multiple scattering effects become larger than the proton scattering.
To incorporate such effects,
we employ the nucleon-target formalism (NTG)~\cite{NTG} as
\begin{align}
  &i\chi(\bm{b})=
  -\int d\bm{r}^P\,\rho^P(\bm{r}^P)\left[1-\exp\left\{-\int d\bm{r}^T\rho^T(\bm{r}^T)\Gamma_{NN}(\bm{s}^P-\bm{s}^T+\bm{b})\right\}\right].
\end{align}
The equation symmetrized for the exchange of the projectile
and target nuclei is employed~\cite{NTG}.
The theory requires the same inputs as these of the OLA and
gives a better description as demonstrated
in Refs.~\cite{Horiuchi06, Horiuchi07, Nagahisa18}.
As a carbon target, we take
the harmonic-oscillator (HO) type density distribution~\cite{Ibrahim09} 
that reproduces the rms point-proton radius
of $^{12}$C, 2.33 fm~\cite{Angeli13}.
As the theory has no adjustable parameter,
the total reaction cross section properly reflects
the profile of the projectile density distribution.
This model was well tested by comparisons of the available experimental
data for neutron-rich unstable
nuclei~\cite{Horiuchi10, Horiuchi12, Horiuchi15, Nagahisa18},
and also used as a standard tool to extract the nuclear matter
radius~\cite{Kanungo11a, Kanungo11b, Bagchi20}.

\section{Results}
\label{results.sec}

\subsection{Blackness of nuclear collisions for medium to heavy nuclei}
\label{blackness.sec}

\begin{figure}[ht]
\begin{center}
  \includegraphics[width=\linewidth]{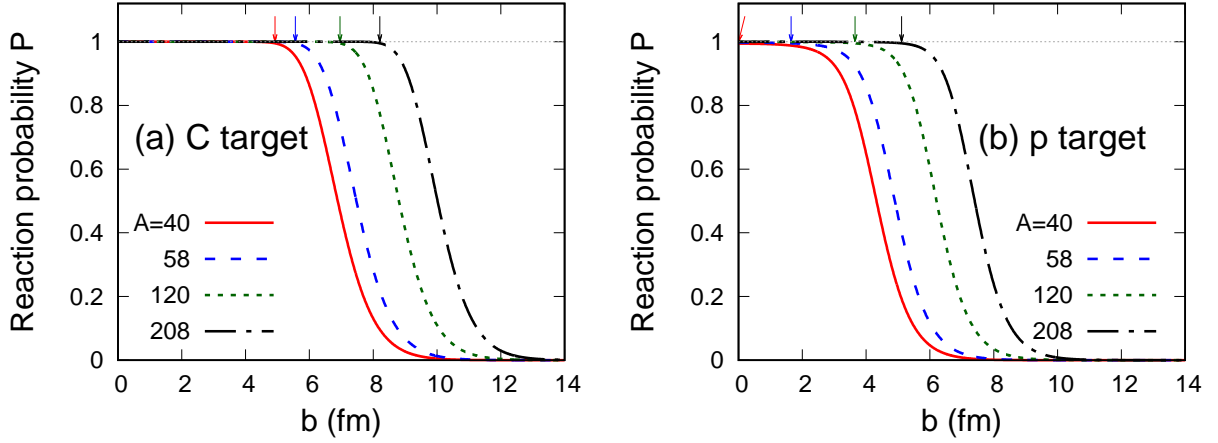}
  \caption{Reaction probabilities of Eq. (\ref{reacprob.eq})
    using 2pF density distributions with $A=40$, 58, 120, and 208
    on (a) carbon and (b) proton targets at 1000 MeV/nucleon.
    The arrows indicate the $S$ values for each system, which are
    defined by the $b$ value where the reaction probability
    becomes $\approx 0.99$.}
  \label{prob2pF0.fig}
\end{center}
\end{figure}

First, we see a global trend of the reaction probabilities
for medium to heavy nuclei. For simplicity,
we employ two-parameter Fermi (2pF) type
density distributions for neutron ($q=n$) and proton ($q=p$)
for a nucleus with the mass number $A$ as
\begin{align}
  \rho_q(r)=\frac{\rho_{0q}}{1+\exp\left[(r-R_0)/a\right]}.
\end{align}
We take a conventional parametrization,
$R_0=1.12A^{1/3}-0.86A^{-1/3}$ and $a=0.54$ fm~\cite{BM}.
The $\rho_{0q}$ value is determined by the normalization condition
$4\pi \int_0^\infty dr\, r^2\rho_{q}(r)=N_q$ with the nucleon number $N_q$
and $A=N_n+N_p$.

Figure~\ref{prob2pF0.fig} compares the reaction probabilities
using the 2pF density distributions with 
$A=40 (N_p=20), 58 (N_p=28), 120 (N_p=50)$ and 208 $(N_p=82)$
on carbon and proton targets at the incident energy of 1000 MeV/nucleon.
For a carbon target, typical behavior of the reaction probabilities is found:
They are unity in the internal regions, where
the two colliding nuclei strongly overlap with each other,
showing the complete absorption in the sense of optics;
and decrease rapidly as less number of nucleons can contribute to the collision
at large impact parameter $b$.
Similar behavior is found for a proton target
but for light projectiles, i.e., $A=40$, the reaction probability
becomes less than unity even at $b=0$.
This is because
the mean-free path $\propto 1/(\rho\sigma^{\rm tot}_{NN})$ of a proton target 
is large enough to penetrate a projectile nucleus,
which is in contrast to the case for a carbon target, where more particles
can contribute to the reaction processes.

Given the properties of the reaction probabilities shown above,
it is convenient to separate the total reaction cross section into
the complete and incomplete absorption parts.
Defining the reaction radius
$R=\sqrt{\sigma_R/\pi}$~\cite{Kohama04, Horiuchi14},
the decomposition reads
\begin{align}
  \sigma_R&=\sigma_{\rm comp}+\sigma_{\rm incomp}
\label{decomrcs.eq}
\end{align}
with
\begin{align}
  \sigma_{\rm comp}&\equiv\pi S^2,\\
        \sigma_{\rm incomp}&\equiv \pi R^2\left[1-\left(\frac{S}{R}\right)^2\right],
\label{decomrcs2.eq}
\end{align}
where the boundary radius $S$ is introduced, which is defined
by the impact parameter that the reaction probability becomes,
say $P(S)\approx 0.99$. 
For $S=R$, i.e., $\sigma_R=\sigma_{\rm comp}$,
it corresponds to the limit that a complete absorption occurs
below a sharp radius $R$, which is simply expressed by
Eq. (\ref{BS.eq}).
The $S$ value can also be zero, i.e., $\sigma_R=\sigma_{\rm incomp}$,
which denotes a ``complete'' incomplete absorption,
where the reaction probability is always less than unity even at $b=0$.
Therefore, $\sigma_{\rm comp}$ and $\sigma_{\rm incomp}$
can respectively be interpreted as black and opaque sphere
parts of the obstacle in optics.
As the complete absorption occurs below the boundary radius $S$,
any structural information in the internal regions cannot
directly be obtained, while
$\sigma_{\rm incomp}$ includes the information on
the density profile near the nuclear surface.
We note, however, that the information on the internal density
  can be obtained indirectly
  by probing the density profile near the nuclear surface,
  see, e.g., for nuclear bubble structure~\cite{Choudhary20}.
We also remark that the earlier study ~\cite{Kox87} proposed
  a similar decomposition, 
  where the energy-independent volume and energy-dependent surface
  terms are assumed, offering
  a good parametrization for $\sigma_R$.
  In the present decomposition,
  to separate the complete and incomplete
  absorption regions based on the reaction probability,
  both the $R$ and $S$ values become energy dependent.

We evaluate the $S$ values for the high energy reactions
on carbon and proton targets with the 2pF projectile density distributions.
The arrows in Fig.~\ref{prob2pF0.fig} indicate these calculated $S$ values,
which are $S=4.9,$ 5.6, 7.0 and 8.2 fm for $A=40, 58, 120$ and $208$,
respectively, for a carbon target, and
$S=0, 1.7, 3.7$ and 5.1 fm for $A=40, 58, 120$ and $208$,
respectively, for a proton target.
The larger the $S$ value, the larger the projectile nucleus becomes, indicating
the penetrability of a probe, i.e., a target nucleus.
This flat behavior is longer for a carbon target
because it involves the size of the target nucleus $\approx$ 3 fm.
Since the reaction probability with $A=40$ for a proton target
is always less than unity, the $S$ value becomes zero by definition,
and hence $\sigma_{\rm comp}=0$.

\begin{figure}[ht]
\begin{center}
  \includegraphics[width=\linewidth]{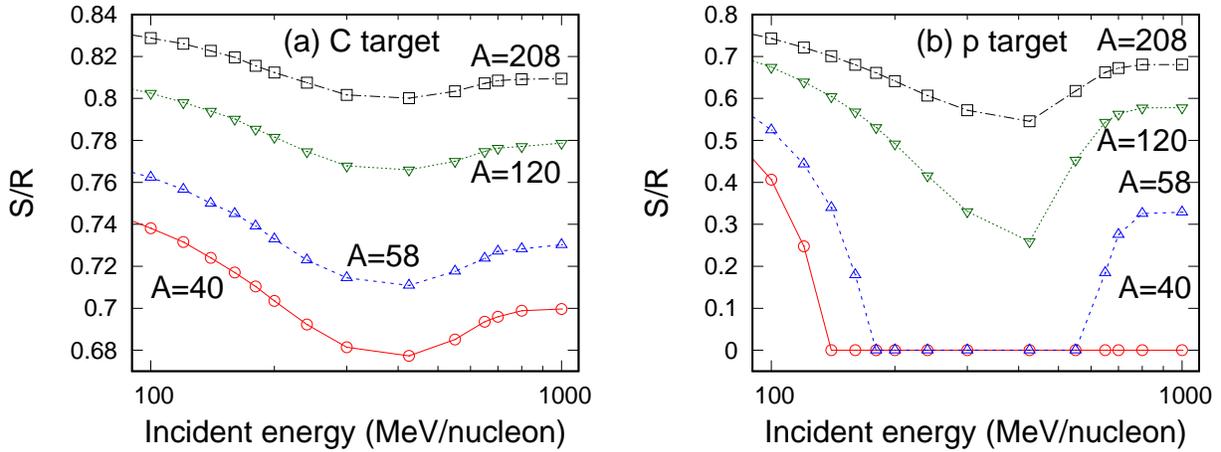}
  \caption{Nuclear blackness, $S/R$, as a function of
    of the incident energy with $A=40$, 58, 120, and 208
    on (a) carbon and (b) proton targets.
    See text for more detail.}
  \label{sigmaA.fig}
\end{center}
\end{figure}

The penetrability of the target nucleus
strongly depends on the magnitudes of
the elementary nucleon-nucleon cross sections.
To see it clearly, we investigate the energy dependence
of the ``blackness'' of the nucleus-nucleus collisions.
Reminding the decomposition of Eqs. (\ref{decomrcs.eq})--(\ref{decomrcs2.eq}),
the ratio $S/R$, which ranges from zero to unity,
can be a measure of the transparency in the colliding processes
and expresses the blackness of the nucleus-nucleus reaction.
Figure~\ref{sigmaA.fig} plots the behavior of
the nuclear blackness $S/R$ as a function of the incident energies
for carbon and proton targets using the 2pF density distributions
with different mass numbers $A=40, 58, 120$, and 208.
The $S/R$ value becomes larger as $A$ increases
because the contribution of the nuclear surface becomes relatively
smaller than that of the nuclear bulk.
For a carbon target, the $S/R$ values are large about 0.6--0.8.
This means that the high-energy nucleus-carbon collision mostly consists
of the complete absorption part.
In the low-incident energy region,
because the elementary nucleon-nucleon cross section is large,
the blackness of the reaction becomes also large.
Here the total reaction cross section
is less sensitive to the internal density of the
projectile nucleus but probes more surface regions.
For a proton target, at the low incident energies $\approx 100$ MeV/nucleon,
the blackness with large $A$ can be comparable to that of the carbon target.
The $S/R$ values are more strongly dependent on the incident energy
since the proton target reflects the elementary cross section
more directly than that of the carbon target~\cite{Horiuchi14,Horiuchi16}.
The blackness becomes minimum for $\approx 200$--600 MeV/nucleon.
where the nucleon-nucleon cross section becomes minimum
and again increases for $\gtrsim 600$ MeV/nucleon,
which reflects the behavior of the nucleus-nucleus total cross sections
included in the profile function.
For $A=40$ and 58, since less number of nucleons contributes
to the reaction processes of a proton target
than these of a carbon target, the nuclear blackness becomes zero,
exhibiting incomplete absorption for a whole impact parameter.

We see that the comparison made here
quantifies the sensitivity of different probes
on the nuclear density profile.
It is interesting to investigate the blackness of the high-energy nuclear
collisions with other probes. See Appendix for the results for
$^4$He and antiproton scattering.

\subsection{Total reaction cross sections involving light nuclei}
\label{carbon.sec}

As seen in the previous subsection,
the nucleus-proton scattering offers a unique opportunity
to exhibit the complete incomplete absorption in the colliding process.
It is interesting to clarify the effect of this incompleteness
on the total reaction cross sections.
Here we investigate proton-rich C isotopes because
some experimental data of the total reaction cross sections
are available for both the carbon and proton targets.
Since the mass number is small, the total reaction cross section
on a proton target can only be written by $\sigma_{\rm incomp}$.

For projectile densities,
we first employ the HO-type density distributions
whose width parameters
are set to reproduce the measured total reaction cross sections
of the proton-rich C isotopes, $^{9-11}$C,
on a carbon target at $\approx$700 MeV/nucleon~\cite{Ozawa95,Ozawa96},
and take commonly for proton and neutron.
The difference between the proton and neutron radii
becomes significant in the proton-rich C isotopes.
To incorpolate this property,
we generate the density distributions
from the Woods-Saxon mean-field as~\cite{BM}
\begin{align}
  V(r)=V_0f(r)+V_1(\bm{l}\cdot\bm{s})r_0^2\frac{1}{r}\frac{df}{dr}(r)
  +V_C(r)\frac{1}{2}(1-\tau_3)
\end{align}
with $\tau_3=1$ for neutron and $-1$ for proton
and $f(r)=\left\{1+\exp[(r-R_{\rm WS})/a_{\rm WS}]\right\}^{-1}$.
We take convenient parametrization given in Ref.~\cite{BM}:
$V_0=-51+33(N-6)/(N+6)\tau_3$ and $V_1=22- 14(N-6)/(N+6)\tau_3$ in units
of MeV with $N$ being the neutron number of C isotopes,
and $V_C$ is the Coulomb potential which only acts on protons.
By applying them to light nuclei,
we use smaller radius and diffuseness parameters
$R_{\rm WS}=1.2(N+6)^{1/3}$ fm and $a_{\rm WS}=0.6$ fm
following Ref.~\cite{Horiuchi21},
and modify the strength of the central potential $V_0$
by multiplying a factor $f$ as $fV_0$ to reproduce
the measured cross sections as was done for the HO density.
The properties of these density distributions are summarized
in Tab.~\ref{C.tab}. For both the HO- and WS-type density distributions,
the behavior of the root-mean-square (rms) matter radius,
$r_m$, is similar: It increases significantly
as going to the proton-rich side mainly
due to the enhancement of the rms proton radius, $r_p$.
Since the present WS parametrization accounts for the isospin dependence,
the difference between the neutron and rms radii,
i.e., the neutron-skin thickness, $r_n-r_p$,
is more significant than the HO one.

\begin{table}[htb]
\centering
\caption{Rms matter, proton, and neutron radii and total reaction
  cross sections of proton-rich C isotopes.
  The HO width and WS potential parameters are respectively adjusted
  to reproduce the interaction cross section data at around 700 MeV/nucleon.
  The cross sections and incident energies are in units of mb and MeV/nucleon,
  respectively.
}
\begin{tabular}{ccccccl}
\hline\hline
     &&$r_p$ (fm)& $r_n$ (fm)& $r_m$ (fm)& $\sigma_R$ (Theo.) & $\sigma_I$ (Expt.) ($E$)\\
\hline
$^9$C& HO&2.58&2.37&2.51&819& $834\pm 18$ ($\sim 720$)~\cite{Ozawa96} \\
     &WS&2.70&1.94&2.48&818& $812\pm 13$ (680)~\cite{Ozawa96}  \\
$^{10}$C&HO&2.36&2.26&2.32&795& $795\pm 12$ ($\sim 720$)~\cite{Ozawa96}\\
     &WS&2.45&2.07&2.30&797&  \\
$^{11}$C&HO&2.18&2.14&2.16&776&$778\pm 40$ ($\sim 730$)~\cite{Ozawa95}\\
     &WS&2.20&2.06&2.13&771& $772\pm 23$ ($700$)~\cite{Ozawa95}\\
\hline\hline
  \end{tabular}
\label{C.tab}
\end{table}

\begin{figure}[ht]
\begin{center}
  \includegraphics[width=0.48\linewidth]{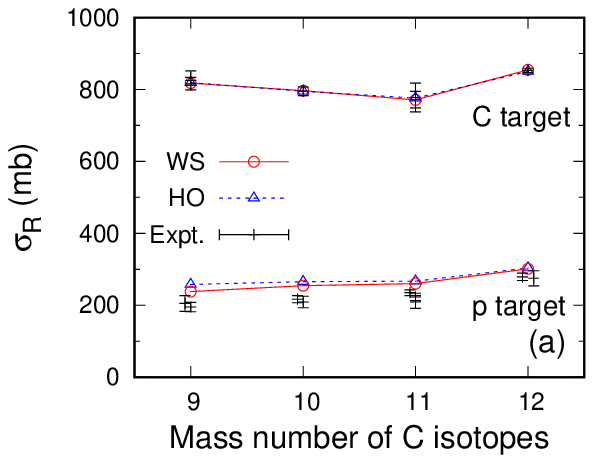}
  \includegraphics[width=0.48\linewidth]{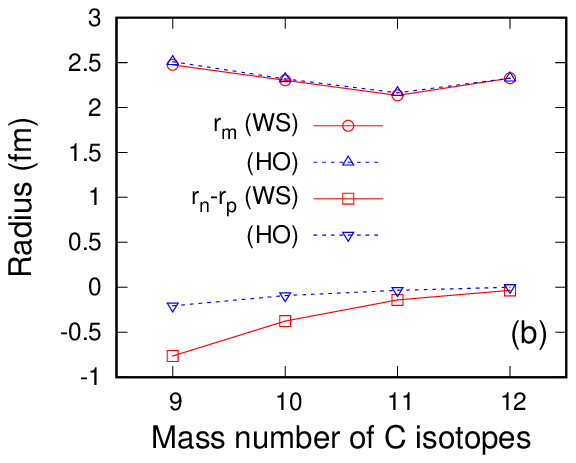}  
  \caption{(a) Total reaction cross sections
    as a function of the mass number of C isotopes.
    Incident energies are 700 and 90 MeV/nucleon for carbon and proton
    targets, respectively.
    The experimental data are taken from
    Refs.~\cite{Ozawa95,Ozawa96,Ozawa01a,Ozawa01b} for a carbon target
    and Refs.~\cite{Auce05,Nishizuka17} for a proton target.
    (b) Rms matter radii $r_m$ and neutron-skin thickness $r_n-r_p$
    as a function of the mass number of C isotopes.
  The WS- and HO-type density distributions are employed.}
  \label{sigmaC.fig}
\end{center}
\end{figure}

Figure~\ref{sigmaC.fig} compares
the total reaction cross sections of the proton-rich C isotopes
on carbon and proton targets.
The rms matter radii and neutron-skin thickness are also plotted.
It appears that the total reaction cross section is well proportional to
the matter radius when a carbon target is used.
However, for a proton target, no increase or even decrease
of the cross section is found for these isotopes
as was seen in Ref.~\cite{Kaki17}.
The calculated cross sections by the WS-type density distributions
better reproduce the experimental data incident at $\approx 90$ MeV
because the WS-type density distributions
have a negative and thicker neutron-skin thickness,
i.e., a positive proton-skin thickness, than that of the HO-type density.
We remark that the thick proton skin structure for $^{9-11}$C
is consistent with the analysis made in Ref.~\cite{Nishizuka17}.
As quantified in Refs.~\cite{Horiuchi14,Horiuchi16},
a thicker proton skin lowers the total reaction cross section
as $\sigma^{\rm tot}_{pp}$ is $\approx 2/3$ of $\sigma^{\rm tot}_{np}$~\cite{PDG,Ibrahim08} at incident energy of $\approx$100 MeV.

\begin{figure}[ht]
\begin{center}
  \includegraphics[width=\linewidth]{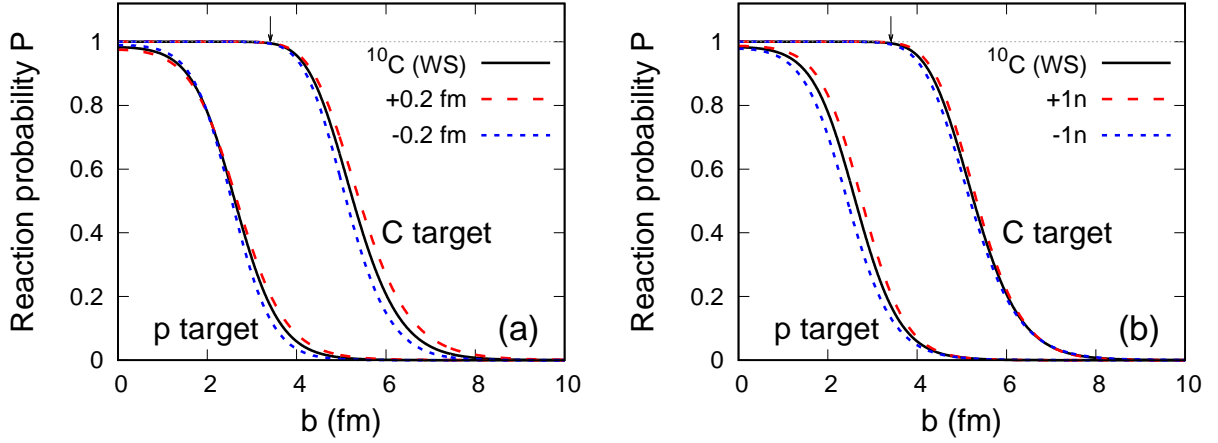}
  \caption{Reaction probabilities of $^{10}$C at 100 MeV/nucleon
    by (a) changing the matter radius
    $\pm 0.2$ fm and (b) removing and adding one neutron while keeping the same matter radius. The arrow indicates the $S$ value of $^{10}$C
    for a carbon target, while $S=0$ for a proton target.}
  \label{prob10C.fig}
\end{center}
\end{figure}

One may ask a question: Why is the different behavior found for the results
with carbon and proton targets?
Figure~\ref{prob10C.fig} displays the reaction probability
of $^{10}$C on carbon and proton targets at 100 MeV/nucleon.
The WS-type density distributions are used.
As expected, for a carbon target,
the reaction probability is unity up to $b=S=3.4$ fm, while
for a proton target, the probability is always
less than unity $(S=0$)
showing the complete incomplete absorption in the whole region. 
For a better understanding,
we investigate the reaction probabilities by changing the
rms matter radius of $^{10}$C by $\pm 0.2$ fm.
The results are drawn in Fig.~\ref{prob10C.fig} (a).
The probabilities appear to change beyond $b>S$:
For a carbon target, they only change in the surface region,
whereas the change in the whole region is found for a proton target as $S=0$.
We also plot in Fig~\ref{prob10C.fig} (b)
the reaction probabilities by changing
the neutron number by $\pm 1$ while keeping the same rms radius.
The behavior is similar to that found in Fig~\ref{prob10C.fig} (a).
Only the probabilities beyond $b>S$ are varied for a carbon target,
while for a proton target, they change in the whole region.

More quantitatively, we evaluate
the difference of the total reaction cross sections
by changing the rms matter radius by $+0.2$ ($-$0.2) fm,
$\Delta\sigma_R$, which are 62 ($-59$) mb
for a carbon and 14 ($-17$) mb for a proton target.
Also, we do similar calculations by adding (removing) one neutron.
They are $19$ ($-21$) mb for a carbon target
and 23 ($-26$) mb for a proton target,
which is similar in both the targets.
To guide the reader, those values are summarized in Tab.~\ref{C2.tab}.
For a carbon target, since the cross section change by
the radius change is larger than that by the mass number change,
the total reaction cross section is in total enhanced
by following the magnitude of the rms matter radius.
On contrary, for a proton target,
the decrease of the cross section due to the change
of the neutron number exceeds the enhancement of the cross section
by the rms matter radius, which can be possible in light nuclei,
where the $\sigma_{\rm incomp}$ contribution is dominant.
This sort of limit of $\sigma_{R}=\sigma_{\rm incomp}$ is
in contrast to the black sphere limit $\sigma_{R}=\sigma_{\rm comp}$.
In this case, the empirical relationship $\sigma_R\propto \pi r_m^2$
appears not to hold because it relies on the assumption that
$\sigma_{\rm incomp}$ is smaller than $\sigma_{\rm comp.}$.

\begin{table}[ht]
  \caption{Modifications of the total reaction cross sections
    with respect to the changes of the density distributions of $^{10}$C
    generated from the WS potential.
        See text for details.}
  \centering
\begin{tabular}{cccccc}
\hline\hline
          & $r_m$ (fm)& $\Delta\sigma_R$(C) & $\Delta\sigma_R$(p)\\
\hline
$^{10}$C &2.30 & --& --&\\
$+0.2$ fm &2.50 & 62  & 14 &\\
$-0.2$ fm &2.10 & $-59$ &$-17$   &\\
$+1n$     &2.30 & 19  &  23 &\\
$-1n$     &2.30 & $-21$  & $-26$   &\\
\hline\hline
  \end{tabular}
\label{C2.tab}
\end{table}

\section{Conclusion}
\label{conclusion.sec}

The total reaction cross section measurement has been a standard tool
to know the nuclear size properties as a basis that
the nucleus-nucleus collision
can be described as optics of strongly absorptive objects.
We have proposed the decomposition of
the total reaction cross section into
complete and incomplete absorption parts
as a measure of the ``blackness'' of the nucleus-nucleus collision.

For a carbon target, the reaction occurs almost completely
when the projectile and target nuclei overlap significantly, that is,
the nuclear collision is ``black'' enough,
showing typical behavior of the cross sections as a function
of the neutron number: The cross section increases
as the nuclear radius increases.
However, it is not always the case
in the nuclear collision involving light elements using a proton target,
where the cross section consists only of the incomplete absorption part.
In that case, the cross section increase due to the enhancement
of the rms matter radius becomes comparable to or even smaller
than the cross section reduction by the mass number change,
and thus the cross section can decrease
despite that the nuclear radius is enhanced towards the proton dripline.

\section*{Acknowledgment}

We thank M. Tanaka for sending us the numerical data
of the experimental cross sections on a proton target.
This work was in part supported by JSPS KAKENHI Grant No. 18K03635.
We acknowledge the Collaborative Research Program 2022, 
Information Initiative Center, Hokkaido University.

\appendix

\section{Blackness of other probes}

\begin{figure}[ht]
\begin{center}
    \includegraphics[width=\linewidth]{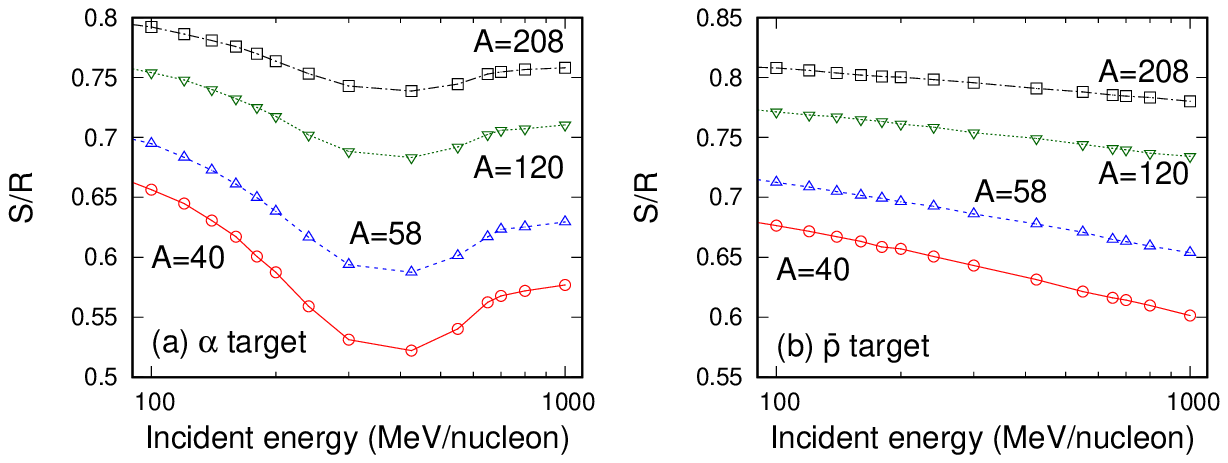}
    \caption{Same as Fig.~\ref{sigmaA.fig} but
      for (a) $\alpha$ and (b) $\bar{p}$
      scattering.}
  \label{sigmaA2.fig}
\end{center}
\end{figure}

The sensitivity of the nuclear density profile on
the total reaction cross sections depends on choices
of a target nucleus and incident energies originating from elementary
hadron-hadron cross sections~\cite{Horiuchi14, Horiuchi16}.
Here we show the blackness of the nuclear scattering by
a $^4$He ($\alpha$) and an antiproton ($\bar{p}$).
Figure~\ref{sigmaA2.fig} plots the $S/R$ values
for $\alpha$ and $\bar{p}$ scattering.
The same 2pF density distributions used in Fig.~\ref{sigmaA.fig}
are employed. We take the HO-type density distribution that reproduces
the charge radius of $^{4}$He~\cite{Angeli13} for $\alpha$-nucleus scattering
and the nucleon-antinucleon ($N\bar{N})$
profile function given in Ref.~\cite{Makiguchi20}
for $\bar{p}$-nucleus scattering.
As seen in Fig.~\ref{sigmaA2.fig} (a),
for the $\alpha$ scattering,
the behavior is almost the same as that of a carbon target
but magnitudes of the $S/R$ values tend to be smaller.
Because the size of $\alpha$ is smaller than that of the carbon target,
the radius where the complete absorption occurs becomes smaller,
and thus the $S$ value is relatively smaller compared to $R$.
For the $\bar{p}$ scattering, as the energy dependence of
$\sigma_{N\bar{N}}^{\rm tot}$ is different from that of $\sigma_{NN}^{\rm tot}$,
the $S/R$ values behave quite differently from the other probes examined
in this paper. It is interesting to note that
typical $S/R$ values are comparable to these of
the $\alpha$ scattering, reflecting large elementary $N\bar{N}$ cross section
and effective interaction range~\cite{Makiguchi20}.

\end{document}